\documentclass{article}

\usepackage{amssymb,amsmath}
\usepackage{epsfig,multirow}

\usepackage{subfigure}

\newcommand\bff{{\bf f}}
\newcommand\bfg{{\bf g}}
\newcommand\bfh{{\bf h}}
\newcommand\x{{\bf x}}
\newcommand\z{{\bf z}}
\newcommand\zero{{\bf 0}}
\newcommand\cc{\mathbb C}

\begin{document}
\title{Accelerating Polynomial Homotopy Continuation
       on a Graphics Processing Unit
       with Double Double and Quad Double Arithmetic}

\author{Jan Verschelde\thanks{email: {\tt jan@math.uic.edu},
        URL: {\tt www.math.uic.edu/}$\sim${\tt jan}}
    ~and~ Xiangcheng Yu\thanks{email: {\tt xiangchengyu@outlook.com}} \\
Department of Mathematics, Statistics, and Computer Science \\
University of Illinois at Chicago, 851 South Morgan (M/C 249) \\
Chicago, IL 60607-7045, USA}

\date{12 June 2015}

\maketitle

\begin{abstract}
Numerical continuation methods
track a solution path defined by a homotopy.
The systems we consider are defined by polynomials in several variables
with complex coefficients.
For larger dimensions and degrees, the numerical conditioning worsens
and hardware double precision becomes often insufficient to reach
the end of the solution path.
With double double and quad double arithmetic, we can solve larger
problems that we could not solve with hardware double arithmetic,
but at a higher computational cost.
This cost overhead can be compensated by acceleration on a Graphics
Processing Unit (GPU).  
We describe our implementation and report on
computational results on benchmark polynomial systems.

\end{abstract}

\section{Introduction}

If we solve linear systems for increasing dimensions,
then we observe increasing numerical condition numbers,
up to the point where double arithmetic no longer suffices
to obtain an accurate solution.
The linear systems we encounter are not random, but originate from
the application of Newton's method to a polynomial system.
For higher precision arithmetic
we use the QD library~\cite{HLB00} and its GPU version~\cite{LHL10}.

Numerical continuation methods apply Newton's method
in a predictor-corrector algorithm to track solution paths,
see~\cite{Li03} for the application to solving polynomial systems.

The focus in this paper is on the acceleration of the tracking
of one single path,
which requires a fine granularity in the parallel algorithm,
because the tracking of one path proceeds step by step,
computing points along a path in a strictly sequential order.
The purpose of our investigations is to find thresholds on
dimensions and sizes of interesting classes of polynomial systems
which clearly benefit from the accelerated path tracking methods.
The goal of acceleration is to offset the
computational overhead caused by the higher precision arithmetic.

In our investigations, we selected the reverse method of algorithmic
differentiation~\cite{GW08} to evaluate the polynomials and their 
derivatives.  Even already for systems of modest size, this method
leads to a good occupancy of the GPU on many blocks of threads.
For fully expanded polynomials with general coefficients,
the cost of evaluation and differentiation agrees with theoretical
complexity bounds~\cite{BS83} for this problem.
For the linear system solving, we have chosen to implement the
modified Gram-Schmidt method instead of row reduction with partial
pivoting, which is more commonly used in path tracking methods.
Because the right-looking algorithm provides more thread-level
parallelism (as pointed out in~\cite{VD08}), 
we first implemented a pure right-looking algorithm.
Combining left and right looking in a tiled
algorithm reduces memory access and yields better speedups.

For very sparse polynomials of low degree, the linear algebra will dominate 
the computational cost of the path tracker.  
For such problems, the number of equations and variables
must be at least several hundreds in order to obtain significant
speedups from the acceleration.
For general polynomials, where the cost of evaluation and
differentiation is substantial, already for systems of modest size
(where the dimension is less than a hundred)
acceleration may already compensate
for the cost of one extra level of precision.

\noindent {\bf Related Work.}
The granularity of parallel homotopy algorithms was investigated
in~\cite{ACW89}.
The consideration in~\cite{ACW89} of fine grained algorithms for 
architectures with fast processor interconnection and relatively 
slow processor speeds is relevant for GPUs.


In computer algebra, the implementation of massively parallel algorithms
for polynomial operations on GPUs are described 
in~\cite{HLMMPX14} and~\cite{MP11}.  
The computation of the Smith normal form 
as needed to solve large systems of binomials (that is: having exactly
two monomials in every equation) using the NVIDIA GTX 780 graphics card
is reported in~\cite{CL14} and in~\cite{CM15}.
The modified Gram-Schmidt method relates to lattice basis reduction
algorithms which have been implemented on graphics cards~\cite{BG12}.

In algorithmic differentiation, reports on parallel implementations 
for multicore architectures can be found in~\cite{BGKW08} 
and on GPUs in~\cite{GPGK08}.

\noindent {\bf Our Contributions.}
In~\cite{VY10}, we experimentally showed that the cost overhead
of double double arithmetic is of a similar magnitude as the
cost of complex double arithmetic and that eight CPU cores
may suffice to offset this overhead in multithreaded implementations.
In~\cite{VY14}, we continued this line of investigation on the GPU,
based on our GPU implementations of evaluation and differentiation
algorithms~\cite{VY12}, combining our GPU implementation of the
the Gram-Schmidt orthogonalization method~\cite{VY13}.
The computational results in~\cite{VY12} and~\cite{VY13} were on
randomly generated data.  The data in this paper comes from
relevant polynomial systems, relevant to actual applications.
In particular, the cyclic $n$-root problems relate to the construction
of complex Hadamard matrices~\cite{Szo11} and the Pieri homotopies solve the
output placement problem in linear systems control~\cite{BB81}.

This paper reports on improvements and the integration of
our building blocks (described in~\cite{VY12, VY13, VY14})
in an accelerated path tracker.
Good speedups relative to the CPU are obtained on benchmark problems,
sufficiently large enough to compensate for the computational overhead
caused by the double double arithmetic.
Several software packages are available to track solution
paths defined by polynomial homotopies, e.g.: Bertini~\cite{Bertini}, 
HOM4PS~\cite{GLL02}, HOM4PS-2.0~\cite{LLT08}, HOM4PS-3~\cite{CLL14},
PHoM~\cite{GKKTFM04}, NAG4M2~\cite{Ley11}, pss~\cite{Mal}, 
and HOMPACK \cite{WBM87, WSMMW97}.
While our work is aimed at accelerating the path trackers in
PHCpack~\cite{Ver99},
also other software packages may benefit from our work.

The focus in this paper is the acceleration of the tracking of one
{\em single} path.  The acceleration of the tracking of {\em many} paths 
defined by a polynomial homotopy is described in~\cite{VY15b}.
The GPU that runs our accelerated path tracker
also hosts our cloud service~\cite{BSVY15}.

\section{Homotopy Continuation}

A family of systems in one parameter $t$ is called a homotopy
and denoted as $\bfh(\x,t) = \zero$.  
The artificial-parameter homotopy we consider is
\begin{equation} \label{eqhomotopy}
   \bfh(\x,t) = \gamma (1-t)^k \bfg(\x) + t^k \bff(\x) = \zero,
\end{equation}
where $\gamma = e^{i \theta}$, $\theta \in [0, 2 \pi)$.
For random~$\gamma$ and with complex arithmetic, the solution
paths are free from singularities for all~$t < 1$,
see for instance~\cite[\S 8.2.1]{SVW05}.
As done in PHCpack~\cite{Ver99}, the relaxation constant
$k \geq 2$ leads to smaller step sizes at the start
when $t \approx 0$ and towards the end when $t \approx 1$.
An input-output specification of the path tracker
summarizes the notations:

\begin{tabbing}
Input: \= $(N, n)$, $N \geq n$: $N$ polynomials in $n$ variables; \\
       \> $\bff(\x) = \zero$, target system, must be solved; \\
       \> $\bfg(\x) = \zero$, start system, with a start solution; \\
       \> $\z \in \cc^n$: $\bfg(\z) = \zero$, a start solution; \\
       \> $(\gamma, k)$ parameters in $\bfh(\x,t) = \zero$
          as in~(\ref{eqhomotopy}). \\
Output: $\z \in \cc^n$: $\bff(\z) = \zero$, a solution of $\bff(\x) = \zero$.
\end{tabbing}

\begin{figure}[hbt]
\begin{center}
\epsfig{figure=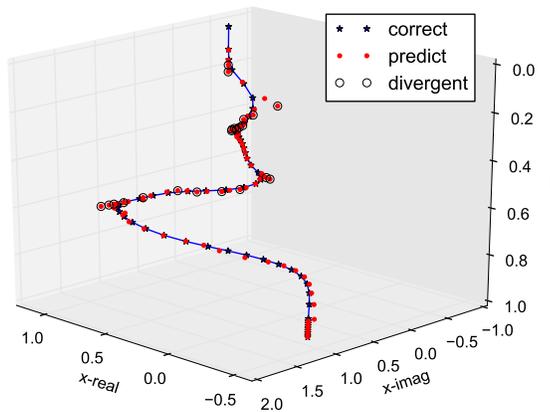,width=8.3cm}
\caption{While tracking one solution path
we show one coordinate of each solution point on the path.
Corrected points (marked by a star) are connected as they lie on the path.
Points that are not connected are predicted points (marked by a dot).
Points from where the corrector diverged are marked by a circle.}
\label{figpathtraker}
\end{center}
\end{figure}

Before we apply Newton's method as a corrector,
a predictor extrapolates the current solution~$\x$
at the current value for~$t$ to the next point on the path.
Based on the outcome of the convergence of Newton's method,
an adaptive step size control strategy increases or decreases
the step size $\Delta t$.

In one Newton step we compute
$\Delta \z$ as the solution of
\begin{equation}
    J_\bff(\z) \Delta \z = - \bff(\z), 
\end{equation}
where $J_\bff$ denotes the matrix of all partial derivatives of
the system $\bff(\x) = \zero$.
Then the current approximation is updated as $\z := \z + \Delta \z$.
To continue with the next Newton step, we run the tests
\begin{equation}
   \| \bff(\z) \| \leq \epsilon_{\| \bff(\z) \|} \quad {\rm and} \quad
   \| \Delta \z \| \leq \epsilon_{\| \Delta \z \|}
\end{equation}
where $\| \bff(\z) \|$ and $\| \Delta \z \|$ are respectively
the backward and forward error of the problem.


Path following in an artificial-parameter homotopy follows
an increment-and-fix strategy.  
The homotopy continuation parameter~$t$ is increased
independently of the coordinates of~$\x$ in the predictor stage, as
\begin{equation}
   t := t + \Delta t, \quad {\rm with} \quad 
   h_{\rm min} \leq \Delta t \leq h_{\rm max}.
\end{equation}

Because the parameter~$t$ in our homotopies~(\ref{eqhomotopy}) 
is artificial we apply the increment-and-fix predictor-corrector methods
to solution paths in complex space.
Our use of extended precision arithmetic is targeted to overcome the 
numerical instabilities that arise from working with higher degree 
polynomials which leads to matrices that may contain numbers of various
magnitudes.  In Table~\ref{tabruncyclic} we report several instances
where one level of precision does not suffice to successfully track
a solution path to the end.

\noindent {\bf Problem Statement.}  The difficulty to accelerate 
the tracking of one single path is that the predictor-corrector
method is a strictly sequential process.
Although we compute many points on a solution path, we cannot compute 
those points in parallel, independently from each other.
In order to move to the next point on the path, the correction for 
the previous point must be completed.
This difficulty requires a fine granularity in the parallel algorithm
and a sufficiently high enough threshold on the dimension of the problem.

\section{Accelerating Path Trackers}

In this section we explain the implementation of path tracking
with predictor-corrector methods in an increment-and-fix approach
for artificial-parameter polynomial homotopies.
We start with the high level descriptions and then go into the
finer details of the accelerated algorithms.

\subsection{Accelerated Predictor-Corrector}

The input format for a polynomial system is a fully expanded
distributed symbolic representation of a list of polynomials.
This representation fits
our choice of polynomial evaluation and differentiation algorithms.
In the reverse mode of algorithmic differentiation, 
the focus is on a product of variables, which provides a large
amount of thread parallelism, suitable for GPUs.
Our path tracker computes with complex arithmetic in three modes:
in double, double double, or quad double precision.

In the first stage, the value for the homotopy continuation parameter
is incremented as $t := t + \Delta t$ and the corresponding solution~$\x$
is updated using extrapolation methods.
Denoting $k$ as the degree of extrapolation polynomial, the values for~$t$ as
$t_0$, $t_1$, $\ldots$, $t_k$, and the corresponding coordinates as
$x_i(t_0)$, $x_i(t_1)$, $\ldots$, $x_i(t_k)$,
then the predicted value for $x_i(t)$ at the new value for~$t$ is
obtained by evaluating a $k$th degree polynomial interpolating
through those points.
As the extrapolation occurs independently for each solution coordinate,
as many threads as the dimension~$n$ are occupied.
But even with complex quad double arithmetic, 
the predictor stage has cost~$O(n)$, 
significantly less than the other stages.

Figure~\ref{figparalgnewton} lists
pseudocode for the accelerated corrector,
illustrating the synchronizations performed by the kernel
launches and the memory transfers from the device to the host to
determine the convergence.

\begin{figure}[hbt]
\begin{center}
\begin{tabbing}
\hspace{6mm} \= Input: \= $Inst$, Polynomial Instructions \\
\> \>  $W$, GPU Workspace \\
\> \>  $P$, parameters for Newton's method \\
\> Out\=put\=: \= $success$ or $fail$ \\
\>    \>   \>  \> updated $W.\x$ \\
\> $last\_max\_eq\_val$ = $P.max\_eq\_val$ \\
\> for $k$ from 1 to $P.max\_iteration$ do \\
\>\>  {\tt GPU\_Eval}($Inst$,$W$)\\
\>\>  laun\=ch kernel {\tt Max\_Array}($W.eq\_val$, $max\_eq\_val$) \\
\>\>     \> with single block \\
\>\>  copy $max\_eq\_val$ from GPU to host \\
\>\>  if ($max\_eq\_val >last\_max\_eq\_val$) \\
\>\>     \>  return $fail$ \\
\>\>  if ($max\_eq\_val < P.tolerance$) \\
\>\>     \>  return $success$\\
\>\>  {\tt GPU\_Modified\_Gram\_Schmidt}($W$)\\
\>\>  launch kernel {\tt Max\_Array}($W.\Delta \x$, $max\_\Delta \x$) \\
\>\>     \> with single block \\
\>\>  copy $max\_\Delta \x$ from GPU to host\\
\>\>  launch kernel {\tt Update\_}$\x$($W.\x$,$W.\Delta \x$)\\
\>\>  if ($max\_\Delta \x < P.tolerance$) \\
\>\>     \>  return $success$\\
\>\>  $last\_max\_eq\_val$ = $max\_eq\_val$ \\
\> return $fail$
\end{tabbing}
\caption{An accelerated Newton's method.}
\label{figparalgnewton}
\end{center}
\end{figure}

The step control coincides with the launching of new kernels
which can be done on the host, as only a constant amount of data
(independent of the size of the problem) needs to be transferred
between host and device.  
In particular, to determine the success of Newton's method,
the device computes a norm of the update $\| \Delta \x \|$,
the norm the residual vector $\| \bff(\x) \|$
and sends this number to the host.  
In reply, the device receives the new step size from the host
before launching the next predictor-corrector kernel.
As the host controls step size, there is a constant amount of
data transfer from device to host.  This size of these data does
not depend on the size of the problem.

Keeping in mind the goal of minimizing the communication between
host and device, we make the following arrangements.
Transferring the polynomial structures, the coefficients, the start
solution from the host to the device occurs only once at the start of 
the path tracking.
In Newton's method, $\| \Delta \x \|$ and $\| \bff(\x) \|$ 
are used to control 
the Newton iteration, sent from device to host.
The host uses these values to check whether the corrector converged
or diverged and to determine whether to launch a new corrector kernel.
The step size $\Delta t$ is controlled by the host.
After each predictor-corrector stage, 
the host updates $t$ by $\Delta t$, and sends this update to the device.
Then the predictor on the device uses the updated values
to predict the next point on the path.
When $t$ gets to $1$, the final solution $\x$ 
is sent from the device to the host.

The computation of the norm of the value of the solution at the system
and the magnitude of the update to the solution can happen in 
ordinary double arithmetic as we are only concerned in the magnitude
of the norms. 
With communications of these 3 double variables between host 
and device, then host can control the device to launch the kernels 
that run the path tracker.

Pseudocode for the path tracker is sketched in Figure~\ref{figparalgpath}.
The {\tt GPU\_Newton} call refers to the algorithm
in Figure~\ref{figparalgnewton}.
The decision to execute the step control at the host is motivated
by its low computational cost and the desire to monitor the quality
of the computations as the tracker advances along a solution path.
Although the latest versions of CUDA allow for separate kernel
launches initiated at the device, the fine granularity of our
algorithms allows for the host to monitor the progress of the
Newton steps.

\begin{figure}[hbt]
\begin{center}
\begin{tabbing}
\hspace{6mm} \= Input: \= $Inst$, Polynomial Instructions \\
\> \>  $W$, GPU Workspace \\
\> \>  $P$, parameters for path tracker \\
\> Out\=put\=: \= $success$ or $fail$ \\
\>    \>   \>  \> $W.\x$, solution for $t=1$ if $success$ \\
\> $t = 0$\\
\> $\Delta t = P.max \Delta t$\\
\> $\#successes = 0$ \\
\> $\#steps = 0$ \\
\>while $t < 1$ do \\
\>\>  if ($\#steps > P.max \#steps$)\\
\>\>\>     return $fail$\\
\>\>  $t = \min(1, t + \Delta t$)\\
\>\>  copy $t$ from host to GPU\\
\>\>  launch kernel {\tt predict}($W.\x\_array$, $W.t\_array$) \\
\>\>  $newton\_success$ = {\tt GPU\_Newton}($Inst$,$W$,$P$)\\
\>\>  if (\=$newton\_success$)\\
\>\>\>   Update pointer of $W.\x$ in $W.\x\_array$\\
\>\>\>   $\#successes$ = $\#successes$ + 1 \\
\>\>\>   if (\= $\#successes > 2$) \\
\>\>\>\>    $\Delta t = \min(2 \Delta t, P.max \Delta t)$ \\
\>\>  else \\
\>\>\>  $\#successes = 0$  \\
\>\>\>  $\Delta t = \Delta t/2$  \\
\>\>  $\#steps = \#steps+1$ \\
\> return $success$
\end{tabbing}
\caption{Accelerated tracking of one single path.}
\label{figparalgpath}
\end{center}
\end{figure}

\subsection{Monomial Evaluation and Differentiation}

We can evaluate and differentiate a single product of variables
(the so-called example of Speelpenning) in two ways:
in the reverse mode or with an arithmetic circuit, 
organized in a binary tree (as we described in~\cite{VY14}).
The latter circuit can use several threads or even occupy an 
entire block to evaluate a single monomial, whereas the straightforward 
application of the reverse mode has to be executed by a single thread.
If $n$ denotes the number of variables in a product,
then the reverse mode writes about $n$ values to global memory
to store intermediate results, whereas the reorganization in a binary tree 
keeps intermediate results in shared memory.  
The drawback however is that the binary tree structure of the
algorithm limits the number of threads that can be used at each level 
of the tree.
The organization in a binary tree is better for memory bound computations,
which is the case when we work in double precision. 
But for compute bound computations, as is the
case with double double and quad double arithmetic, 
the reverse mode is better.

To balance the work loads for each block, we sort the products of the
variables according to the number of variables that appear in the product,
so all threads in a single block work on products of similar size.
For improved access to global memory, the data structures are designed
to allow for coalesced reading and writing.
The instructions to evaluate a product include the
number of variables that occur in the product 
and the positions to indicate where each variable occurs in the product.
Aligning numbers of variables and indices to positions,
all threads are reading from global memory from consecutive locations
at the same step.  Access to global memory for intermediate and
final results is handled in the same way.
Table~\ref{tabevaldiffprod} illustrates the evaluation and
differentiation of a product of variables.

\begin{figure}[hbt]
\begin{center}
\epsfig{figure=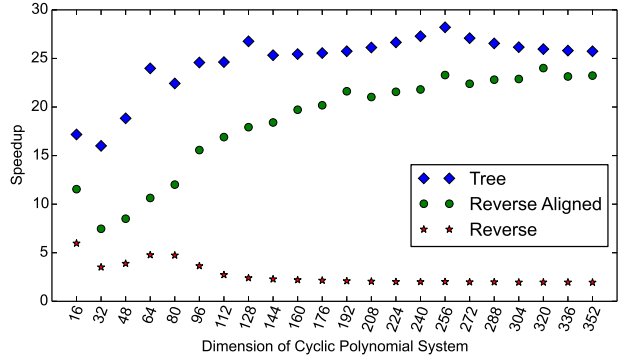,width=8.3 cm}
\caption{Monomial evaluation and differentiation speedup comparison 
in complex double precision}
\label{figmonspeedup}
\end{center}
\end{figure}

Figure~\ref{figmonspeedup}  illustrates the comparison of the original 
reverse mode, reverse mode with aligned memory and the tree mode in 
complex double precision.
The polynomial system in this experiment is defined in~(\ref{eqcyclicsys}).
After alignment, the reverse mode improves, 
especially for larger dimension systems.
The tree mode still works better for complex double precision. 
But for larger monomials, due to the limits of threads in the tree mode, 
reverse mode with aligned memory almost catches up.
This new reverse mode
also works better for double double and quad double arithmetic.

\begin{table}[hbt]
{\small
\begin{center}
\setlength{\tabcolsep}{3pt}
    \begin{tabular}{ | c | c | c | c | c |}
    \hline
     tidx & 0 & 1 & 2 & 3 \\ \hline
   $m_{tidx}$ & $x_0x_1x_2$ & $x_3x_4x_5$ & $x_2x_3x_4x_5$ & $x_0x_1x_3x_4x_5$
 \\ \hline
   \multirow{5}{*}{$\displaystyle \frac{\partial m_{tidx}}{\partial x_j}$}   &  &  &  & \\ 
      & $x_0$    & $x_3$    & $x_2$        & $x_1$ \\ 
      & $x_0 \star x_1$ & $x_3 \star x_4$ & $x_2 \star x_3$   
 & $x_1 \star x_2$  \\ 
      &          &          & $x_2x_3 \star x_4$  & $x_1x_2 \star x_3$ \\ 
      &          &          &              & $x_1x_2x_3 \star x_4$ \\ \hline
     \multicolumn{5}{c}{the forward calculation, from the top to bottom row}
    \end{tabular}

    \begin{tabular}{ | c | c | c | c | c |}
    \hline
   tidx & 0 & 1 & 2 & 3 \\ \hline
   $m_{tidx}$ & $x_0x_1x_2$ & $x_3x_4x_5$ & $x_2x_3x_4x_5$ & $x_0x_1x_3x_4x_5$
 \\ \hline
    \multirow{5}{*}{$\displaystyle \frac{\partial m_{tidx}}{\partial x_j}$} 
  & $x_1 \star x_2$ & $x_3 \star x_4x_5$ & $x_3 \star x_4x_5$ 
& $x_2 \star x_3x_4x_5$ \\ 
 & $x_0 \star x_2$    & $x_3 \star x_5$& $x_2 \star (x_4 \star x_5)$
       & $x_1 \star (x_3 \star x_4x_5)$  \\ 
 & $x_0x_1$ & $x_3x_4$ & $x_2x_3 \star x_5$ & $x_1x_2 \star (x_4 \star x_5)$   \\ 
 &          &          & $x_2x_3x_4$  & $x_1x_2x_3 \star (x_5)$  \\ 
 &          &          &              & $x_1x_2x_3x_4$\\ \hline
   \multicolumn{5}{c}{backward and cross products, from bottom to top row}
    \end{tabular}
\caption{Evaluating four monomials
 $x_0x_1x_2$, $x_3x_4x_5$, $x_2x_3x_4x_5$, $x_0x_1x_3x_4x_5$.
 The $\mbox{tidx}$  in the tables below stands for thread index.}
\label{tabevaldiffprod}
\end{center}
}
\end{table}

\subsection{Number of Threads in the Summation}

We improved the efficiency of our summation routines over those 
in~\cite{VY14}. We reorganized the summation algorithm 
so many threads work on one sum.
This led to an increased amount of parallelism and an improved
coalesced access to the main memory of the device.
Even already in double precision arithmetic, double digits speedups
of the accelerated code were obtained.

Tables~\ref{tabsumcyc128} 
and~\ref{tabsumcyc352} illustrate that the best speedups
are obtained if four threads are collaborating 
to perform the summation.
Working with a finer granularity increase the parallelism,
but also leads to a larger number of blocks.
Observe that on the GPU all times in the quad double column 
are all less than the time on the CPU.
With acceleration, we can quadruple the precision
and still be about twice faster than without acceleration.
The polynomial system in this experiment is defined
in~(\ref{eqcyclicsys}).

\begin{table}[t]
\begin{center}
\setlength{\tabcolsep}{4pt}
{\small
\begin{tabular}{r|r|rr|rr|rr}
    & \#Th & {\tt D}~~ & {\tt S}~~ 
           & {\tt  DD}~~ & {\tt S}~~ & {\tt QD}~~ & {\tt S}~~   \\ \hline
CPU &      & 3.86 &       &  21.67 &       & 136.64 &        \\ \hline
GPU &   1  & 0.65 &  5.90 &   1.26 & 17.22 &   1.89 &  72.29 \\ 
    &   2  & 0.44 &  8.85 &   0.56 & 38.57 &   1.38 &  99.18 \\ 
    &   4  & 0.34 & 11.35 &   0.53 & 41.21 &   1.35 & 100.90 \\
    &   8  & 0.37 & 10.53 &   0.63 & 34.39 &   1.54 &  88.44 \\
    &  16  & 0.45 &  8.56 &   0.86 & 25.29 &   1.97 &  69.20 \\
    &  32  & 0.65 &  5.89 &   0.96 & 22.59 &   2.58 &  53.00 \\
\end{tabular}
}
\caption{Summation times in milliseconds 
for cyclic 128-roots with an increasing number of threads ${\tt \#Th}$
per monomial on the GPU, for complex arithmetic in
double (${\tt D}$), double double (${\tt DD}$),
and quad double precision (${\tt QD}$),
with Speedup (${\tt S}$) relative to one CPU core.}
\label{tabsumcyc128}
\end{center}
\end{table}

\begin{table}[hbt]
\begin{center}
\setlength{\tabcolsep}{4pt}
{\small
\begin{tabular}{r|r|rr|rr|rr}
    & {\tt \#Th} & {\tt D}~~ & {\tt S}~~  & {\tt DD}~~ & {\tt S}~~ 
    & {\tt QD}~~ & {\tt S}~~  \\ \hline
CPU &      & 95.02 &       & 422.71 &       & 2512.66 &        \\ \hline
GPU &   1  & 13.10 &  7.25 &  20.23 & 20.90 &   32.36 &  77.64 \\
    &   2  &  7.50 & 12.66 &   9.97 & 42.41 &   24.61 & 102.10 \\
    &   4  &  6.32 & 15.04 &   9.69 & 43.63 &   24.65 & 101.92 \\
    &   8  &  6.91 & 13.75 &  11.24 & 37.61 &   27.60 &  91.03 \\
    &  16  &  8.37 & 11.36 &  15.06 & 28.06 &   33.36 &  75.32 \\
    &  32  & 12.37 &  7.68 &  16.35 & 25.86 &   39.37 &  63.82 \\
\end{tabular}
}
\caption{Summation times in milliseconds 
for cyclic 352-roots with an increasing number of threads ${\tt \#Th}$
per monomial on the GPU, for complex arithmetic in
double (${\tt D}$), double double (${\tt DD}$),
and quad double precision (${\tt QD}$).
with Speedup (${\tt S}$) relative to one CPU core.}
\label{tabsumcyc352}
\end{center}
\end{table}

\begin{table}[h]
\begin{center}
{\small
\setlength{\tabcolsep}{4pt}
\begin{tabular}{r|r|rr|rr|rr}
    & {\tt \#Th} & {\tt D}~~ & {\tt S}~~  
    & {\tt DD}~~ & {\tt S}~~ & {\tt QD}~~ & {\tt S}~~    \\ \hline
CPU &      & 196.90 &       &  918.25 &       & 3145.17 &         \\ \hline
GPU &   1  &  77.61 &  2.54 &  152.48 &  6.02 &  241.87 &   13.00 \\
    &   2  &  29.43 &  6.69 &   46.45 & 19.77 &  124.68 &   25.23 \\
    &   4  &  18.19 & 10.83 &   27.09 & 33.89 &   71.72 &   43.86 \\
    &   8  &  14.16 & 13.90 &   18.96 & 48.44 &   46.51 &   67.63 \\
    &  16  &  12.76 & 15.43 &   15.30 & 60.02 &   35.20 &   89.34 \\
    &  32  &  14.58 & 13.50 &   14.46 & 63.52 &   30.53 &  103.02 \\
\end{tabular}
}
\caption{Summation times in milliseconds 
for the Pieri problem of dimension~103 
with an increasing number of threads {\tt \#Th} 
per monomial on the GPU, for complex arithmetic in
double ({\tt D}), double double ({\tt DD}),
and quad double precision ({\tt QD}).
with Speedup ({\tt S}) relative to one CPU core.}
\label{tabsumpieri103}
\end{center}
\end{table}

Table~\ref{tabsumpieri103} displays timings to evaluate and
differentiation a polynomial system arising from the minor expansions,
as in~(\ref{eqdetexp}).
Increasing the number of collaborating threads keeps increasing the
speedup.  Except for the case when only one thread works on one monomial,
the speedup is sufficiently large for the overhead caused by
quadrupling the precision.

\section{Test Problems}

We chose two classes of benchmark
polynomial systems that can be formulated for any dimension.
In the first problem, we bootstrap from a linear system into a gradually
higher dimensional and higher degree problem.
Monodromy is applied in the second benchmark problem
and the homotopies connect polynomial systems of the same complexity.

\subsection{Matrix Completion with Pieri Homotopies}

Our first class of test problems has its origin in the output pole
placement problem in the control of linear systems.
We may view this problem as an inverse eigenvalue problem~\cite{KRW04}.
The polynomial equations arise from minor expansions on
\begin{equation} \label{eqdetexp}
   \det(A | X) = 0, \quad A \in \cc^{n \times m}, 
\end{equation}
and where $X$ is an $n$-by-$p$ matrix ($m + p = n$) of unknowns.
For example, a 2-plane in complex 4-space (or equivalently,
a line in projective 3-space) is represented as
\begin{equation} \label{eqxmatrix}
   X = 
   \left[
      \begin{array}{cc}
         1 & 0 \\
         x_{2,1} & 1 \\
         x_{2,2} & x_{3,2} \\
         0 & x_{4,2} \\
      \end{array}
   \right].
\end{equation}
To determine for the four unknowns in~$X$ we need four equations
as in~(\ref{eqdetexp}), which via expansion results in four 
quadratic equations.
In the application of Pieri homotopy algorithm
(see e.g.~\cite{HSS98}, \cite{HV00}, \cite{LWW02},
and~\cite{Sot97}), we consider matrices $X$:
\begin{equation}
   \left[
      \begin{array}{cc}
         1 & 0 \\
         0 & 1 \\
         0 & x_{3,2} \\
         0 & 0 \\
      \end{array}
   \right],
   \quad
   \left[
      \begin{array}{cc}
         1 & 0 \\
         0 & 1 \\
         0 & x_{3,2} \\
         0 & x_{4,2} \\
      \end{array}
   \right],
   \quad
   \left[
      \begin{array}{cc}
         1 & 0 \\
         x_{2,1} & 1 \\
         0 & x_{3,2} \\
         0 & x_{4,2} \\
      \end{array}
   \right],
\end{equation}
and then ends in the matrix X of~(\ref{eqxmatrix}).
Each matrix in the sequence introduces one new variable
and the homotopy starts at a solution of the previous homotopy,
extended with a zero value for the new variable,
each time a new matrix~$A$ is introduced.

Using a superscript to index a sequence of matrices,
$A^{(i)} \in \cc^{n \times m}$, $i=1,2,\ldots,k$,
Pieri homotopies are defined as
\begin{equation}
   \bfh(\x,t) = 
   \left\{
      \begin{array}{l}
         \det(A^{(i)} | X) = 0, ~ i=1,2,\ldots,k-1, \\
         \det(t A^{(k)} + (1-t) S_X | X) = 0, 
      \end{array}
   \right.
\end{equation}
where $S_X$ is a special matrix which ensures that for $t = 0$,
we have start solutions by setting the bottommost variables of~$X$ to zero.
Because of the similarities in the monomial structure, for this
fully determined type of Pieri homotopy we may consider
as last equation in the homotopy
\begin{equation}
        t \det(A^{(k)} | X) + (1-t) \det(S_X | X) = 0.
\end{equation}
In the sequence of homotopies, the index~$k$ runs from 1 to $m \times p$.
Because in our setup, we track one single path, we may start at $k = m-1$,
which corresponds to a linear system as only the last column of~$X$
contains variables.  As $k$ increases, the polynomial homotopy becomes
more and more nonlinear.  In the last stages of the homotopy, for $p = 3$,
the cost of evaluation via the minor expansions becomes cubic in~$n$.
As the cost of evaluation and differentiation becomes dominant, 
the most important factor lies in the summation of
the many terms in every polynomial.

\subsection{Monodromy on Cyclic $n$-roots}

Our second test problem is the cyclic $n$-roots problem, 
denoted by $\bff(\x) = \zero$,
$\bff = (f_1, f_2, \ldots, f_n)$, with

\begin{equation} \label{eqcyclicsys}
     \begin{array}{l}
        f_1 = x_{0}+x_{1}+ \cdots +x_{n-1}, \\
        f_2 = x_{0}x_{1}+x_{1}x_{2}+ \dots +x_{n-2}x_{n-1}+x_{n-1}x_{0}, \\
        f_i = \displaystyle\sum_{j=0}^{n-1} ~ \prod_{k=j}^{j+i-1}
        x_{k~{\rm mod}~n}, i=3, 4, \ldots, n-1, \\
        f_n = x_{0}x_{1}x_{2} \cdots x_{n-1} - 1. \\
     \end{array}
\end{equation}
Observe the increase of the degrees: $\deg(f_k) = k$.
This implies that evaluating the system at points with
coordinates of modulus different from one has the potential
to lead to either very small or to very large numbers.
\begin{figure*}[ht]
 \centering
 \subfigure[Path 1]{
 \includegraphics[width=4.1cm]{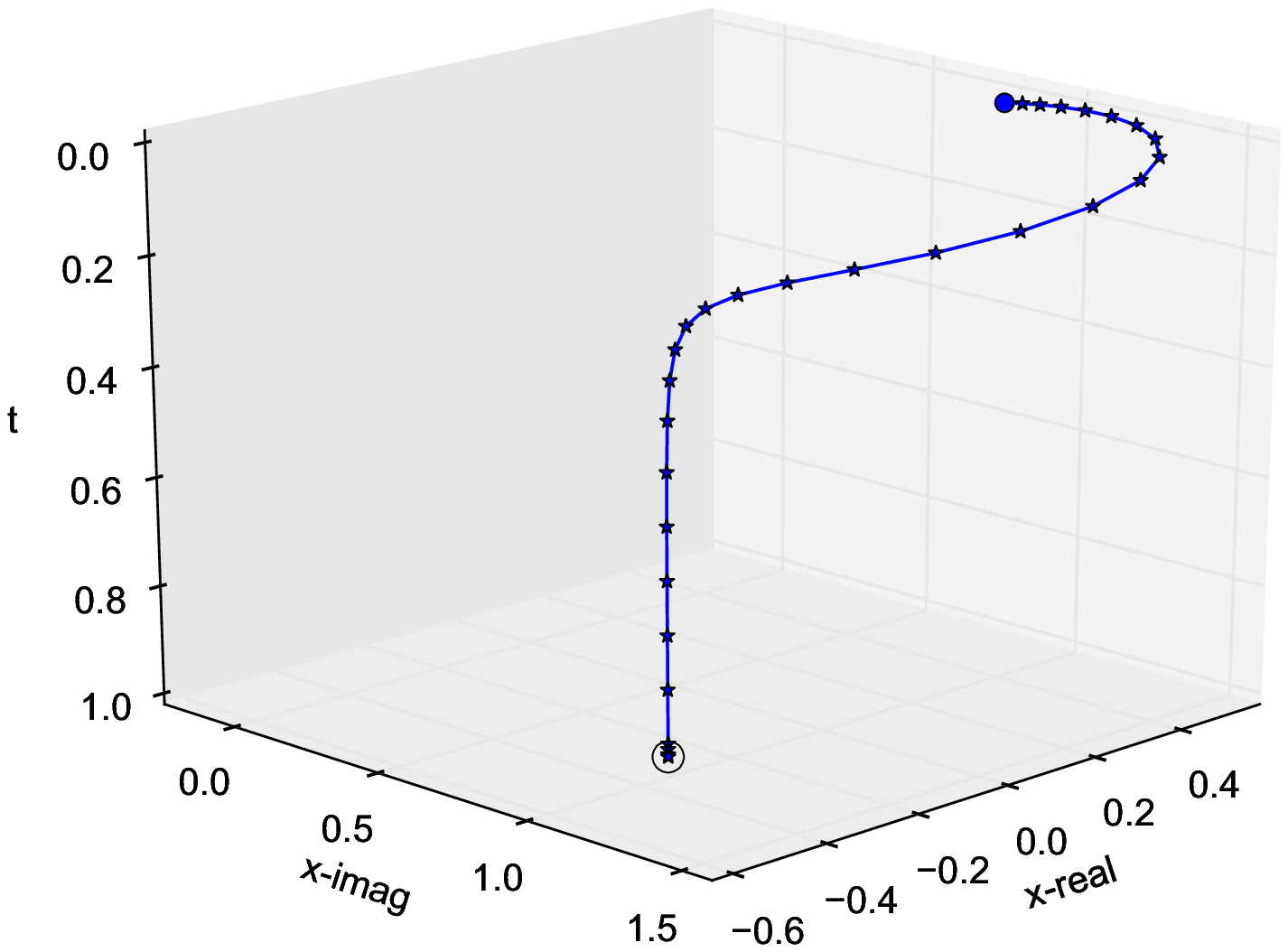}
}
 \subfigure[Path 2]{
 \includegraphics[width=4.1cm]{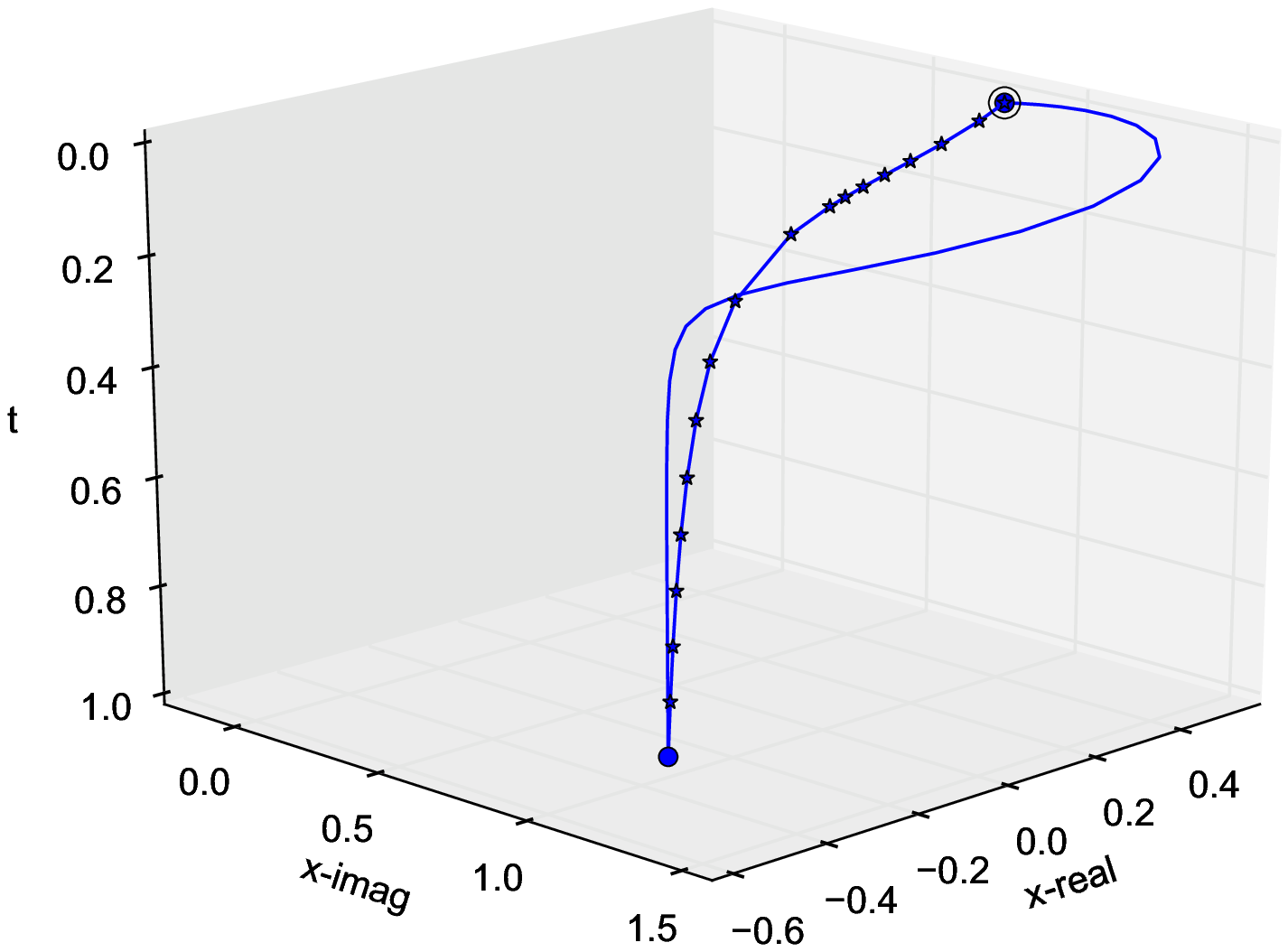}
}
 \subfigure[Path 3]{
 \includegraphics[width=4.1cm]{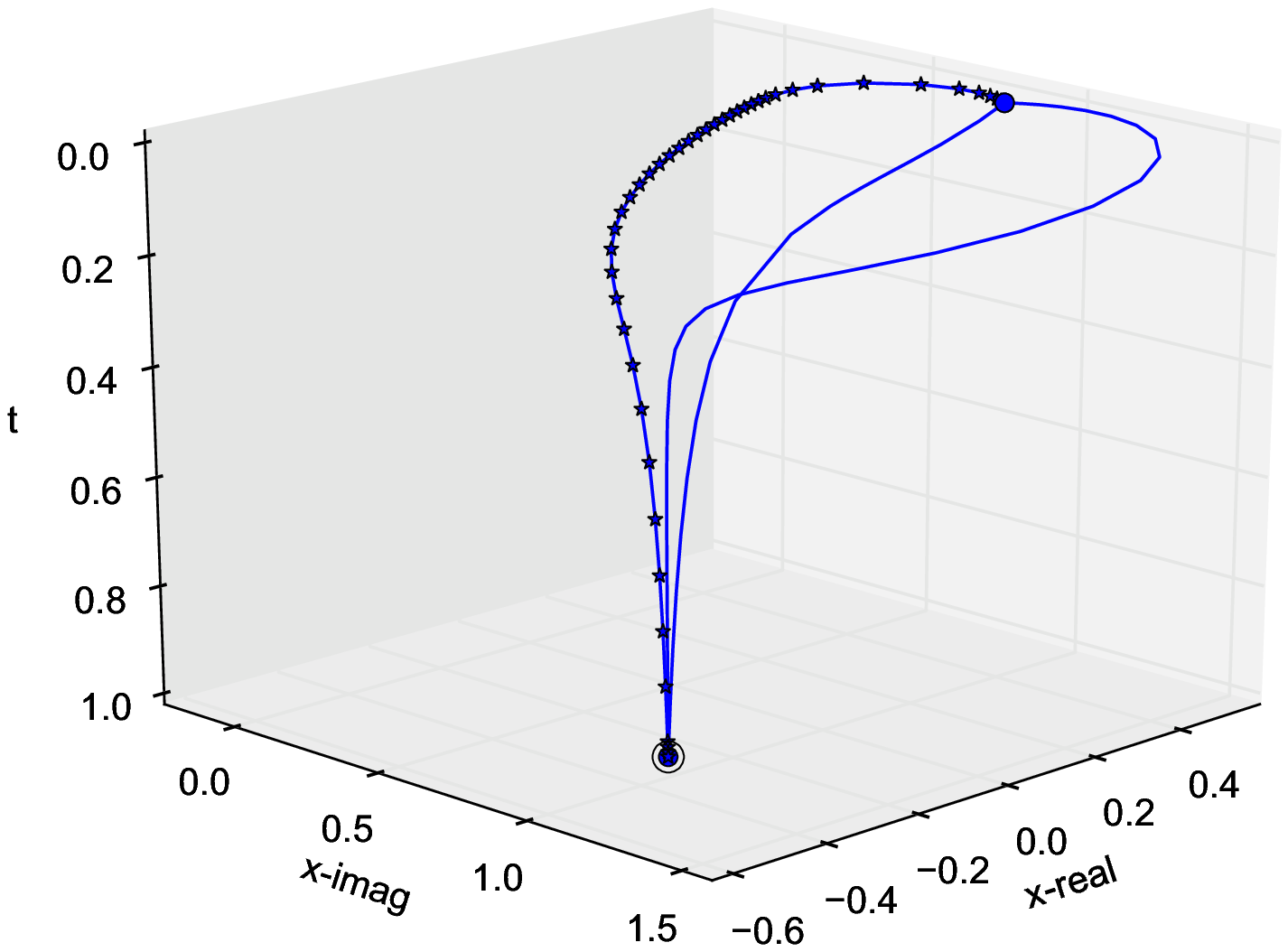}
}
 \subfigure[Path 4]{
 \includegraphics[width=4.1cm]{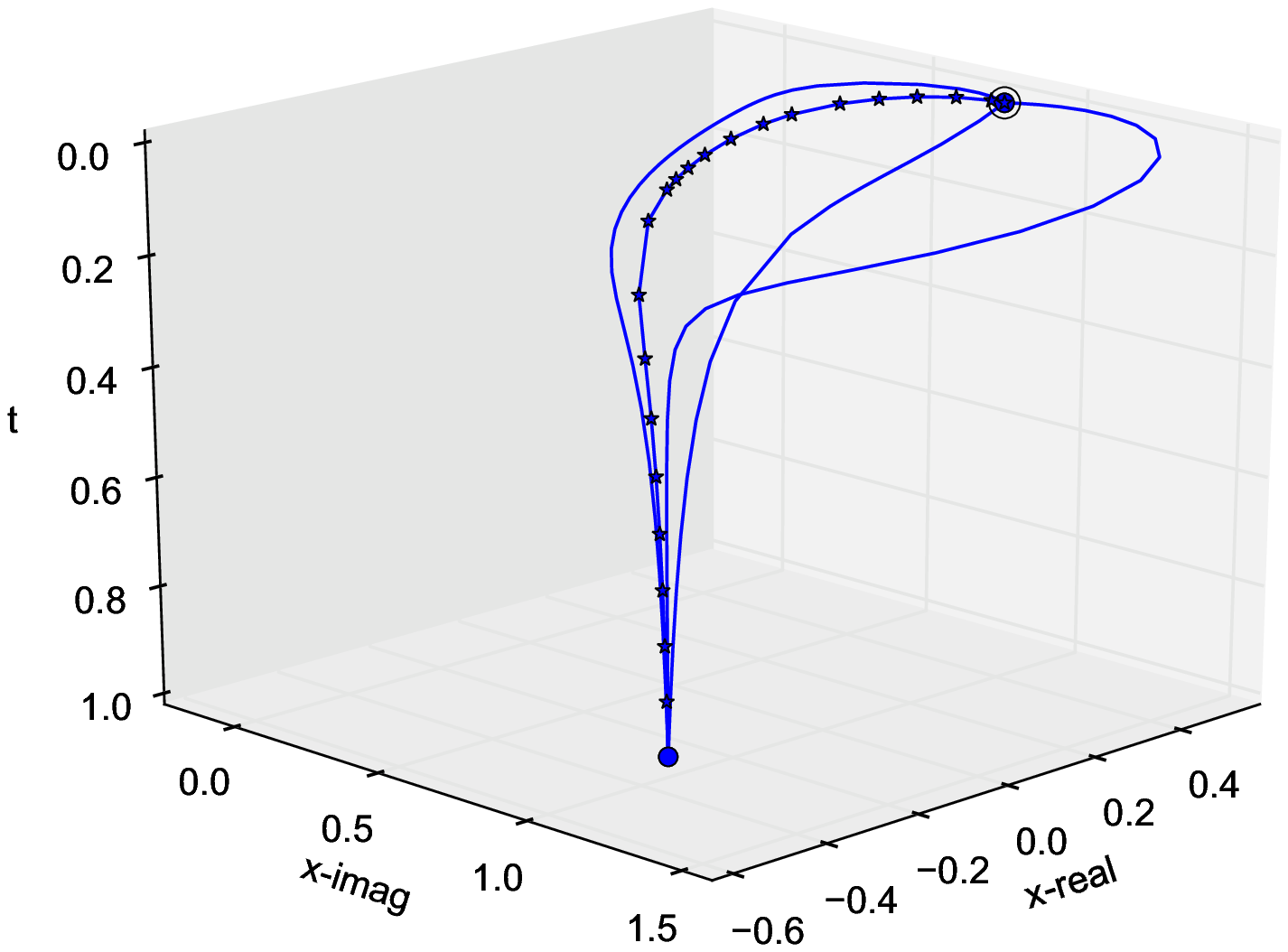}
}
 \subfigure[Path 5]{
 \includegraphics[width=4.1cm]{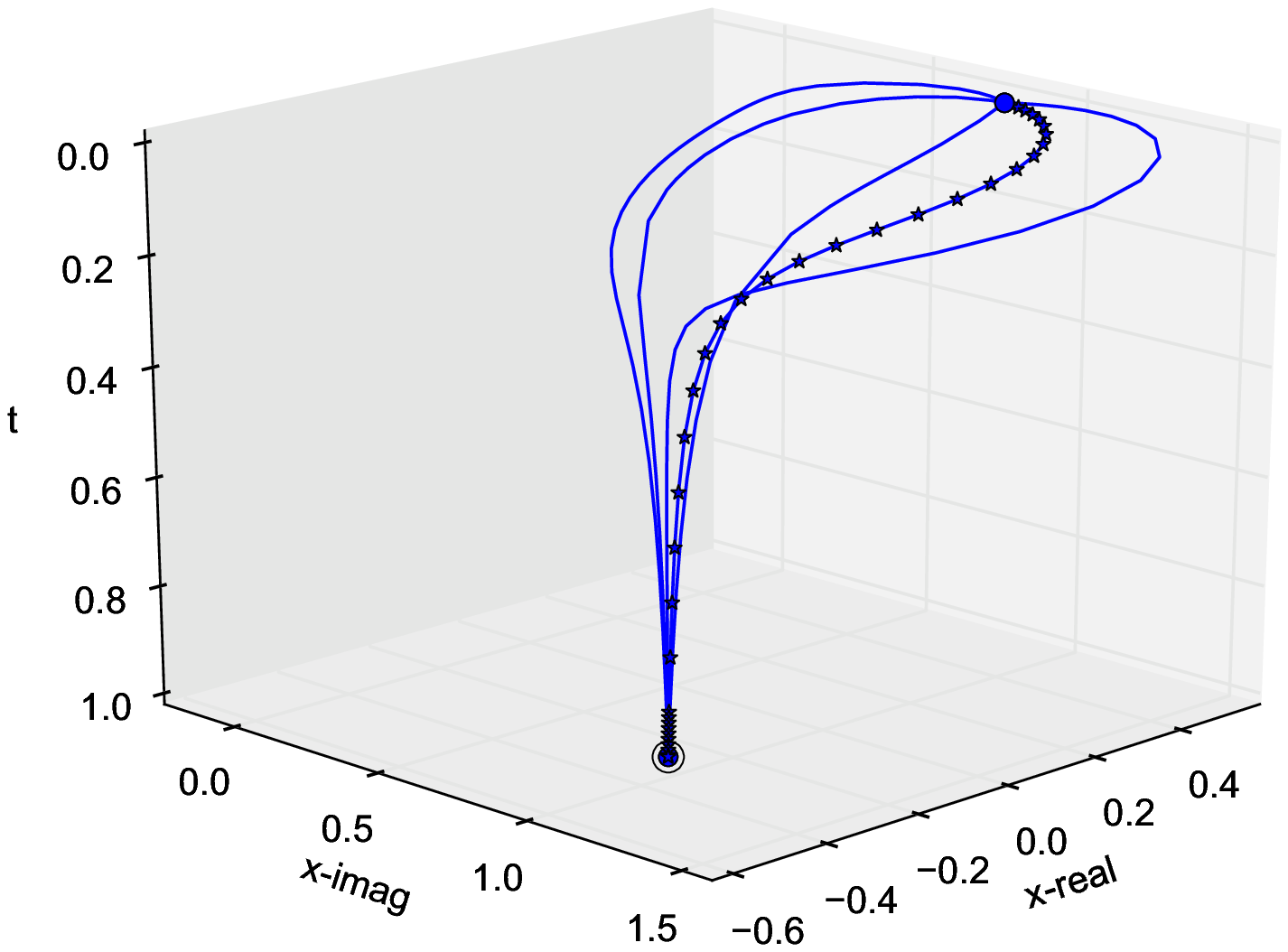}
}
 \subfigure[Path 6]{
 \includegraphics[width=4.1cm]{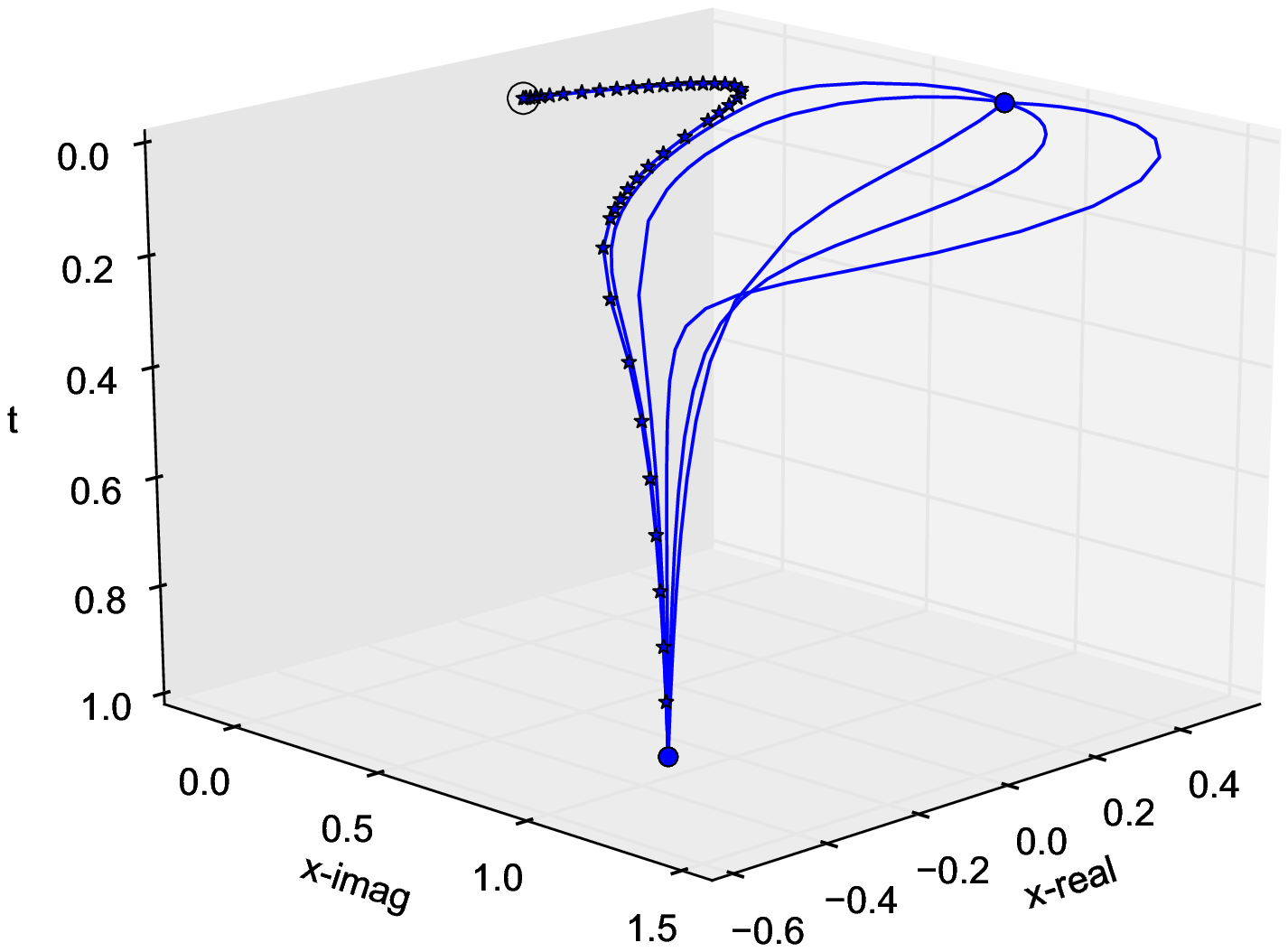}
}
 \subfigure[Path 7]{
 \includegraphics[width=4.1cm]{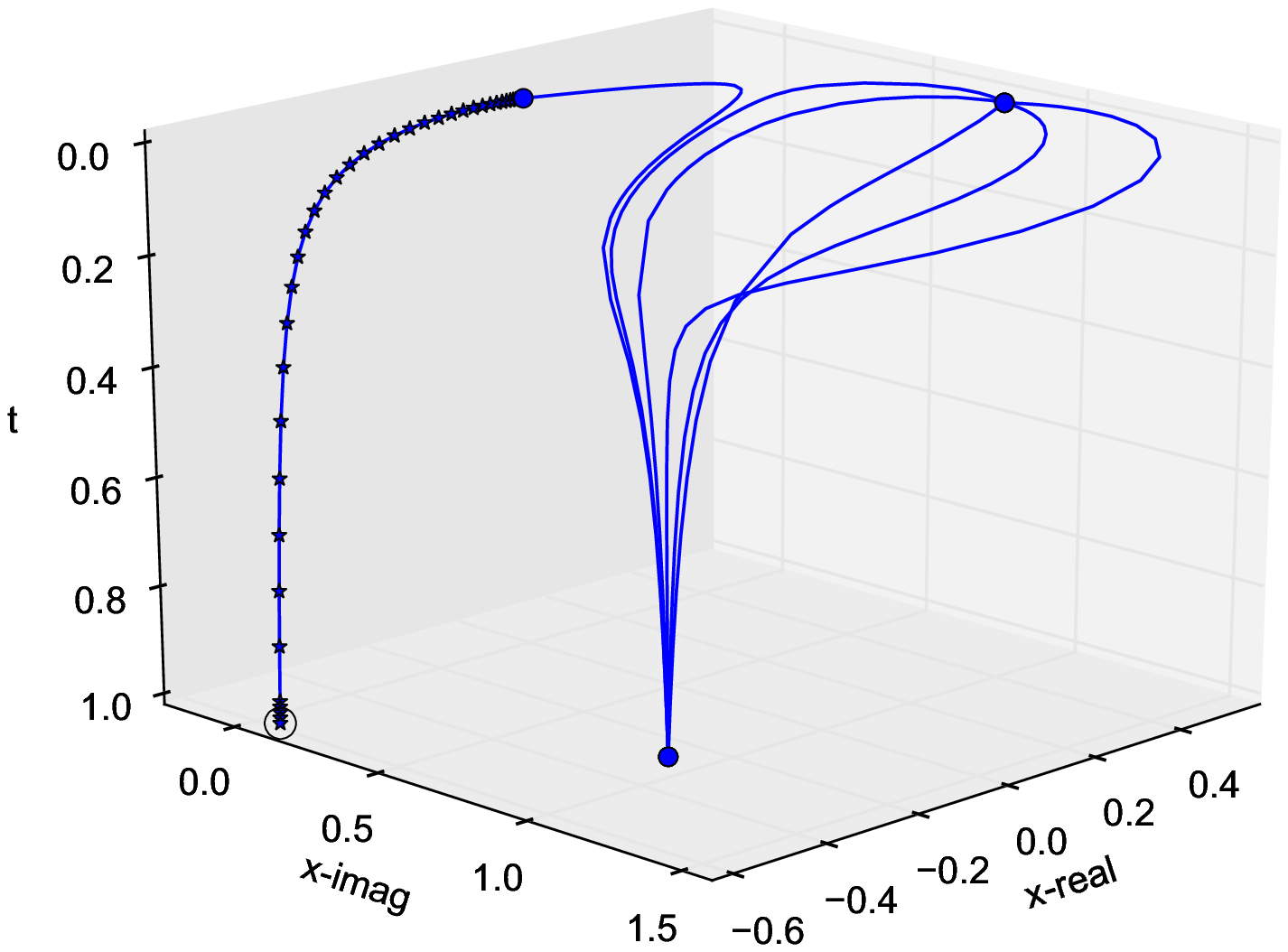}
}
 \subfigure[Path 8]{
 \includegraphics[width=4.1cm]{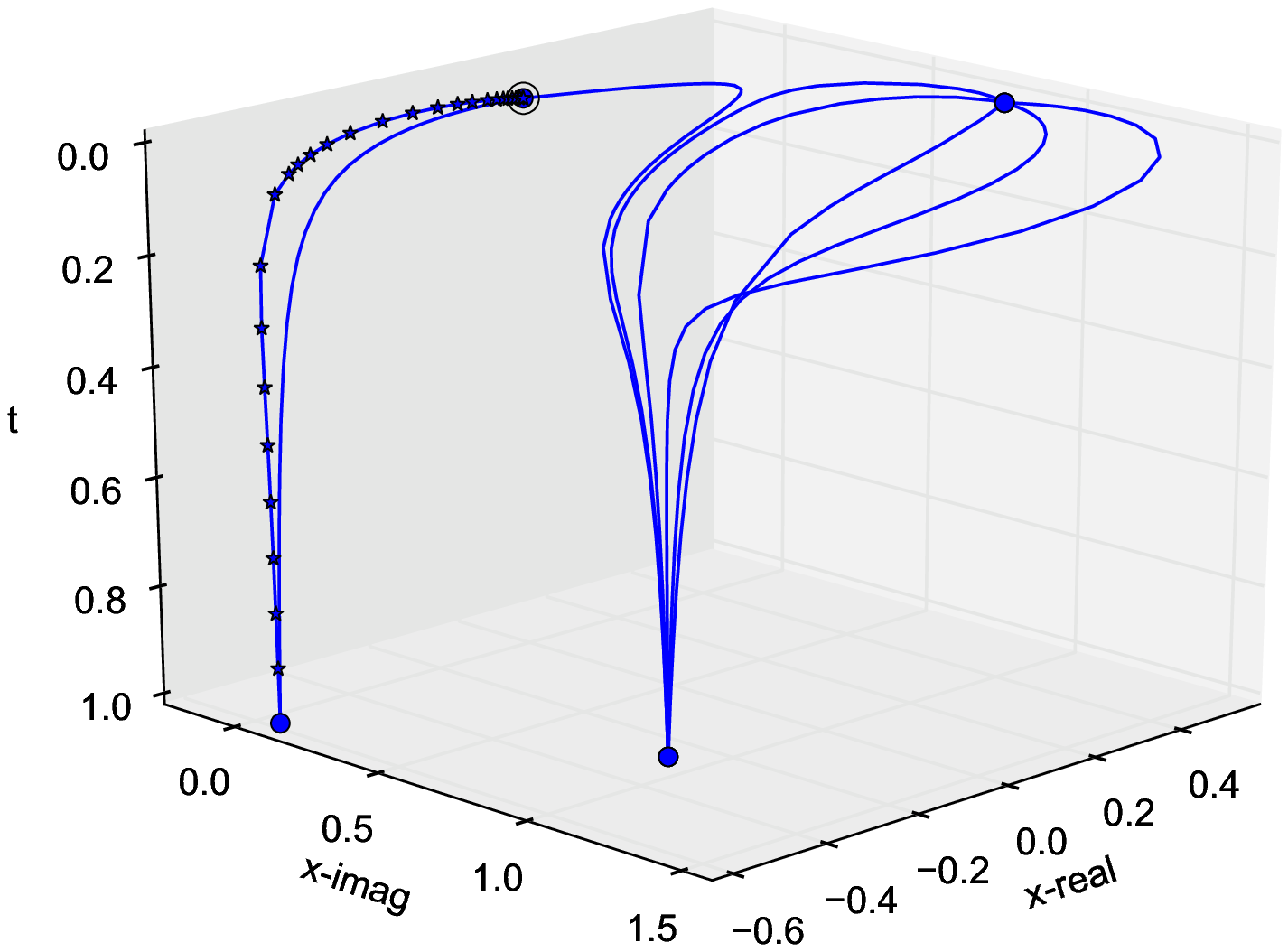}
}
 \caption{Visualization of monodromy on cyclic 4-roots, which
has two solution sets of degree~2.
Each subfigure adds a new path.
Corrected points on path are marked by a star 
and the target point is circled.}
\label{fig:figmonodromy}
\end{figure*}

Backelin's Lemma~\cite{Bac89} states that this system has
a solution set of dimension~$m-1$ for~$n = \ell m^2$,
where $\ell$ is no multiple of $k^2$, for $k \geq 2$.
The system benchmarks polynomial solvers, 
see e.g.~\cite{DKK03}, \cite{Fau01}, and~\cite{Sab11}.
In~\cite{AV13} we derived an explicit parameter representation
for those positive dimensional cyclic $n$-roots solution sets.
To compute the degree of the sets, we add as many linear
equations $\bf L$ (with random complex coefficients)
as the dimension of the set and count the number of
solutions of the system $\bff(\x) = \zero$, augmented with $\bf L$:
\begin{equation} \label{eqaugsys}
   \left\{
      \begin{array}{l}
         \bff(\x) = \zero \\
         {\bf L}(\x) = \zero.
      \end{array}
   \right.
\end{equation}
The explicit representation of the cyclic $n$-roots solution sets
allows for a quick calculation of the degrees,
displayed in Table~\ref{tabcycdeg}.
From~\cite[Proposition~4.2]{AV12}, we have that the degree $d = m$
for $n = m^2$ and this result extends for $n = \ell m^2$.

\begin{table}[hbt]
{\small
\begin{center}
\setlength{\tabcolsep}{4pt}
\begin{tabular}{c|rrrrrrrrrr}
  $n$ & 16 & 32 & 48 & 64 & 80 & 96 & 128 & 144 & 160 & 176 \\ \hline
  $d$ &  4 &  4 &  4 &  8 &  4 &  4 &   8 &  12 &   4 &   4 \\ \hline \hline
  $n$ & 192 & 208 & 240 & 256 & 272 & 288 & 304 & 320 & 336 & 352 \\ \hline
  $d$ &   8 &   4 &   4 &  16 &   4 &  12 &   4 &   8 &   4 &   4
\end{tabular}
\caption{Degrees $d$ of the cyclic $n$-roots solution sets.}
\label{tabcycdeg}
\end{center}
}
\end{table}

\noindent Observe that many solution sets in Table~\ref{tabcycdeg}
have degree four.  A fourth-order predictor will give accurate
predictions on a surface of degree four.
Therefore, the numerically harder problems are those dimensions for
which the degree of the solution set is larger than four.
For cyclic 64-roots double precision is no longer sufficient. 

As done in PHCpack~\cite{SVW03}, with monodromy, the degree is computed
numerically, using a sequence of homotopies:
\begin{eqnarray}
   \bfh_\alpha(\x,t) & = &
   \left\{
      \begin{array}{c}
         \bff(\x) = \zero \\
         \alpha (1-t) {\bf L}(\x) + t {\bf K}(\x) = \zero
      \end{array}
   \right. \\
   \bfh_\beta(\x,t) & = &
   \left\{
      \begin{array}{c}
         \bff(\x) = \zero \\
         \beta (1-t) {\bf K}(\x) + t {\bf L}(\x) = \zero
      \end{array}
   \right.
\end{eqnarray}
where ${\bf K}(\x) = \zero$ is as ${\bf L}(\x) = \zero$ another set of 
linear equations with random coefficients
and where $\alpha$ and $\beta$ are different random complex constants.
One loop consists in tracking one path defined by
$\bfh_\alpha(\x,t) = \zero$ and $\bfh_\beta(\x,t) = \zero$.
In both cases $t$ goes from~0 to~1.
See Figure~\ref{fig:figmonodromy}.

After sufficiently many loops, each time for different values
of the random constants~$\alpha$ and~$\beta$, we will find as many
different solutions of the system~(\ref{eqaugsys}) as the degree
of the solution set, as in Table~\ref{tabcycdeg}.

\section{Computational Results}

In this section we report timings and speedups.

\subsection{Hardware and Software Environments}

We implemented the path tracker with the gcc compiler
and version 6.5 of the CUDA Toolkit.
Our NVIDIA Tesla K20C,
which has 2496 cores with a clock speed of 706~MHz,
is hosted by a Red Hat Enterprise Linux workstation of Microway,
with Intel Xeon E5-2670 processors at 2.6~GHz.
Our code is compiled with the optimization flag {\tt -O2}.

The code is at
{\tt https://github.com/yxc2015/Path}.
The benchmark data were prepared 
with {\tt phcpy}~\cite{Ver14},
the Python interface to PHCpack, in particular, with the scripts
{\tt backelin.py} and {\tt pierisystems.py} of the {\tt examples}
directory of the {\tt PHCpy} package.

\subsection{Running Pieri Homotopies}

Tables~\ref{tabpieri1} and~\ref{tabpieri2}
summarize the execution of a sequence of Pieri homotopies, 
for dimensions $n$ ranging from 32 and 96.
For each path, we list the number $m$ of predictor-corrector stages,
which is the same for the host and device, as we use the same
step length control strategy for both.

Note that timings in Tables~\ref{tabpieri1} and~\ref{tabpieri2}
are rounded.  The speedups are computed on the original timings.

\begin{table}[hbt]
\begin{center}
{\small
\setlength{\tabcolsep}{4pt}
\begin{tabular}{rrrrr|rrrrr}
 $n$ & $m$ & cpu & gpu & S~~ & $n$ & $m$ & cpu & gpu & {\tt S}~~ \\ \hline
 32 &  25 &   0.05 &   0.05 &    1.1 & 65 &  63 &    5.6 &    0.7 &    8.0 \\ 
 33 & 150 &   0.70 &   0.55 &    1.3 & 66 & 187 &   25.3 &    2.4 &   10.7 \\ 
 34 &  65 &   0.44 &   0.30 &    1.5 & 67 &  98 &   17.4 &    1.3 &   13.1 \\ 
 35 & 101 &   0.84 &   0.49 &    1.7 & 68 & 239 &   50.0 &    3.2 &   15.8 \\ 
 36 &  36 &   0.40 &   0.21 &    1.9 & 69 & 244 &   68.0 &    3.7 &   18.4 \\ 
 37 &  10 &   0.13 &   0.06 &    2.1 & 70 & 118 &   40.4 &    2.0 &   20.1 \\ 
 38 &  37 &   0.56 &   0.25 &    2.3 & 71 &  41 &   17.0 &    0.8 &   21.9 \\ 
 39 &  24 &   0.44 &   0.18 &    2.4 & 72 &  89 &   41.8 &    1.8 &   23.2 \\ 
 40 &  19 &   0.39 &   0.15 &    2.6 & 73 &  99 &   44.2 &    1.8 &   24.6 \\ 
 41 &  52 &   1.12 &   0.41 &    2.7 & 74 &  85 &   41.9 &    1.7 &   25.0 \\ 
 42 &  66 &   1.38 &   0.48 &    2.9 & 75 &  89 &   50.3 &    1.9 &   26.0 \\ 
 43 &  72 &   1.67 &   0.55 &    3.0 & 76 & 246 &  136.0 &    4.6 &   29.8 \\ 
 44 &  23 &   0.61 &   0.19 &    3.2 & 77 & 100 &   53.3 &    1.9 &   27.7 \\ 
 45 &  16 &   0.45 &   0.13 &    3.4 & 78 &  81 &   45.7 &    1.6 &   28.2 \\ 
 46 &  25 &   0.74 &   0.21 &    3.5 & 79 & 272 &  210.0 &    6.2 &   34.0 \\ 
 47 &  27 &   0.90 &   0.24 &    3.7 & 80 & 226 &  158.0 &    5.5 &   28.7 \\ 
 48 &  53 &   1.69 &   0.45 &    3.8 & 81 &  50 &   39.0 &    1.3 &   29.4 \\ 
 49 &  32 &   1.05 &   0.27 &    3.9 & 82 & 116 &   91.2 &    3.1 &   29.1 \\ 
 50 & 108 &   3.77 &   0.94 &    4.0 & 83 & 136 &  107.2 &    3.6 &   29.5 \\ 
 51 &  48 &   1.67 &   0.41 &    4.1 & 84 &  69 &   59.2 &    2.0 &   29.6 \\ 
 52 &  79 &   2.97 &   0.71 &    4.2 & 85 & 248 &  206.5 &    6.9 &   30.1 \\ 
 53 &  53 &   2.06 &   0.47 &    4.4 & 86 & 181 &  166.5 &    5.3 &   31.2 \\ 
 54 &  91 &   3.37 &   0.75 &    4.5 & 87 &  32 &   31.1 &    1.0 &   30.5 \\ 
 55 &  18 &   0.90 &   0.19 &    4.6 & 88 &  36 &   37.3 &    1.2 &   30.2 \\ 
 56 &  28 &   1.37 &   0.29 &    4.7 & 89 &  94 &  113.7 &    3.2 &   36.0 \\ 
 57 &  45 &   2.01 &   0.42 &    4.7 & 90 &  73 &   75.7 &    2.5 &   29.9 \\ 
 58 &  34 &   1.69 &   0.35 &    4.8 & 91 &  66 &   68.4 &    2.3 &   30.1 \\ 
 59 &  29 &   1.41 &   0.28 &    5.0 & 92 &  90 &   98.0 &    3.2 &   30.3 \\ 
 60 & 111 &   5.70 &   1.13 &    5.1 & 93 & 102 &  112.6 &    3.7 &   30.3 \\ 
 61 &  67 &   3.77 &   0.74 &    5.1 & 94 &  41 &   40.2 &    1.3 &   30.4 \\ 
 62 &  42 &   2.40 &   0.46 &    5.3 & 95 &  53 &   61.4 &    1.8 &   34.4 \\ 
 63 &  85 &   4.85 &   0.91 &    5.3 & 96 &  64 &   67.2 &    2.2 &   30.8 \\ 
 64 &  63 &   3.36 &   0.62 &    5.4 \\
\end{tabular}
}
\caption{Running the first instance of Pieri homotopies in 
complex double double arithmetic, from dimensions 32 to 96.
The units of the times on CPU and GPU are seconds.
The last column for each dimension contains the speedup~{\tt S}.}
\label{tabpieri1}
\end{center}
\end{table}

\begin{table}[hbt]
\begin{center}
{\small
\setlength{\tabcolsep}{4pt}
\begin{tabular}{rrrrr|rrrrr}
 $n$ & $m$ & cpu & gpu & S~~ & $n$ & $m$ & cpu & gpu & {\tt S}~~ \\ \hline
 32 &  16 &   0.03 &   0.03 &    1.1 &  68 &   51 &   11.3 &    0.7 &   15.6 \\ 
 33 &  68 &   0.38 &   0.30 &    1.3 &  69 &   60 &   17.9 &    1.0 &   18.2 \\ 
 34 &  21 &   0.16 &   0.10 &    1.5 &  70 &   22 &    7.7 &    0.4 &   21.1 \\ 
 35 &  19 &   0.20 &   0.11 &    1.7 &  71 &  156 &   62.0 &    2.8 &   22.1 \\ 
 36 &  18 &   0.20 &   0.11 &    1.9 &  72 &   39 &   19.4 &    0.8 &   23.4 \\ 
 37 &  34 &   0.44 &   0.21 &    2.1 &  73 &   49 &   26.9 &    1.1 &   24.2 \\ 
 38 &  32 &   0.44 &   0.19 &    2.3 &  74 &   98 &   56.7 &    2.3 &   24.6 \\ 
 39 &  41 &   0.65 &   0.26 &    2.5 &  75 &   74 &   43.6 &    1.7 &   25.9 \\ 
 40 &  12 &   0.25 &   0.09 &    2.6 &  76 &   63 &   38.8 &    1.5 &   26.4 \\ 
 41 &  17 &   0.36 &   0.13 &    2.9 &  77 &   37 &   27.1 &    1.0 &   27.0 \\ 
 42 &  29 &   0.65 &   0.23 &    2.9 &  78 &   95 &   67.9 &    2.5 &   27.6 \\ 
 43 &  66 &   1.47 &   0.48 &    3.1 &  79 &  112 &   76.6 &    2.8 &   27.7 \\ 
 44 &  10 &   0.28 &   0.08 &    3.2 &  80 &  157 &  115.0 &    4.0 &   28.8 \\ 
 45 &  46 &   1.30 &   0.39 &    3.4 &  81 &  321 &  265.2 &    7.6 &   35.0 \\ 
 46 &  31 &   0.85 &   0.24 &    3.5 &  82 &   63 &   47.9 &    1.6 &   29.6 \\ 
 47 &  51 &   1.60 &   0.44 &    3.6 &  83 &   42 &   33.3 &    1.1 &   29.9 \\ 
 48 &  16 &   0.54 &   0.14 &    3.8 &  84 &   19 &   17.3 &    0.6 &   30.0 \\ 
 49 &  16 &   0.58 &   0.15 &    3.9 &  85 &  224 &  188.1 &    6.2 &   30.2 \\ 
 50 &  24 &   0.91 &   0.23 &    3.9 &  86 &  159 &  147.9 &    4.3 &   34.1 \\ 
 51 &  62 &   2.31 &   0.56 &    4.1 &  87 &  252 &  199.4 &    6.4 &   31.0 \\ 
 52 &  40 &   1.52 &   0.36 &    4.3 &  88 &  574 &  431.2 &   13.3 &   32.4 \\ 
 53 &  46 &   2.09 &   0.49 &    4.2 &  89 &  213 &  171.3 &    5.5 &   30.9 \\ 
 54 &  33 &   1.62 &   0.37 &    4.4 &  90 &  137 &  129.9 &    3.6 &   35.9 \\ 
 55 &  79 &   3.84 &   0.86 &    4.5 &  91 &  187 &  157.8 &    5.0 &   31.7 \\ 
 56 &  36 &   1.71 &   0.36 &    4.7 &  92 &  250 &  219.8 &    6.1 &   36.2 \\ 
 57 &  42 &   2.23 &   0.48 &    4.6 &  93 &  847 &  646.1 &   19.3 &   33.4 \\ 
 58 &  29 &   1.58 &   0.33 &    4.8 &  94 &  199 &  169.1 &    4.9 &   34.6 \\ 
 59 &  37 &   1.98 &   0.40 &    4.9 &  95 &  108 &   96.1 &    3.1 &   31.0 \\ 
 60 &  16 &   0.95 &   0.19 &    5.0 &  96 &  190 &  230.8 &    6.2 &   37.1 \\ 
 61 &  37 &   2.10 &   0.41 &    5.2 &  97 &  161 &  305.3 &    6.7 &   45.8 \\ 
 62 &  50 &   2.97 &   0.57 &    5.2 &  98 &   76 &  264.7 &    4.4 &   60.5 \\ 
 63 &  34 &   1.95 &   0.36 &    5.4 &  99 &   75 &  322.6 &    5.5 &   58.7 \\ 
 64 &  75 &   4.54 &   0.84 &    5.4 & 100 &  242 & 1367.2 &   23.0 &   59.4 \\ 
 65 &  83 &   7.93 &   0.96 &    8.3 & 101 &  809 & 4655.7 &   70.7 &   65.9 \\ 
 66 & 195 &  27.20 &   2.56 &   10.6 & 102 & 1016 & 5231.7 &   75.6 &   69.2 \\ 
 67 & 154 &  26.43 &   2.07 &   12.8 & 103 &  375 & 2923.2 &   44.0 &   66.5 \\ 
\end{tabular}
}
\caption{Running the second instance of Pieri homotopies in 
complex double double arithmetic, from dimensions 32 to 103.
The units of the times on CPU and GPU are seconds.
The last column for each dimension contains the speedup {\tt S}.}
\label{tabpieri2}
\end{center}
\end{table}

\begin{figure}[hbt]
\begin{center}
\epsfig{figure=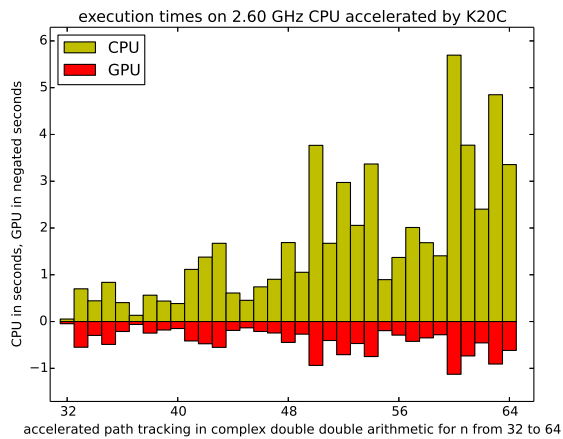,width=8.3cm}
\caption{Visualization of the first half of Table~\ref{tabpieri1}.
While the times on the GPU are always smaller than the CPU times for
the same dimension, fluctuations in the number of steps, e.g.:
compare $n=59$ with $n=60$ may matter more.}
\label{figpieri1plothalf}
\end{center}
\end{figure}

\begin{figure}[hbt]
\begin{center}
\epsfig{figure=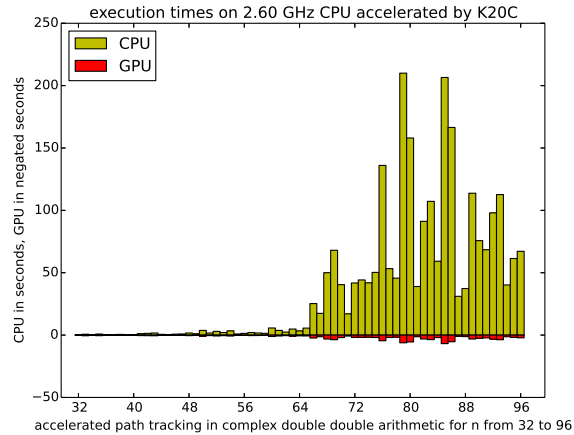,width=8.3cm}
\caption{Visualization of the entire Table~\ref{tabpieri1}.
With a more than double speedup of the GPU over the CPU,
fluctuations in times caused by a variable number of steps
no longer matter.}
\label{figpieri1plotall}
\end{center}
\end{figure}

Tables~\ref{tabpieri1} and~\ref{tabpieri2} are visualized in
Figures~\ref{figpieri1plothalf} and~\ref{figpieri1plotall}.
Because the fluctuations in the number of predictor-corrector steps
along a path can vary by a factor as large as five, the single digit
speedups obtained by acceleration in low dimensions is often in the
same range as the factor in the fluctuations of the timings.
While fluctuations in larger dimensions remain of the same order,
the double digit speedups make that with acceleration we may increase the
dimension, compare for example the lines for $n = 63$ and $n = 96$
in Table~\ref{tabpieri1} and still be faster: 2.2 seconds versus 4.85
seconds.

\begin{figure}[hbt]
\begin{center}
\epsfig{figure=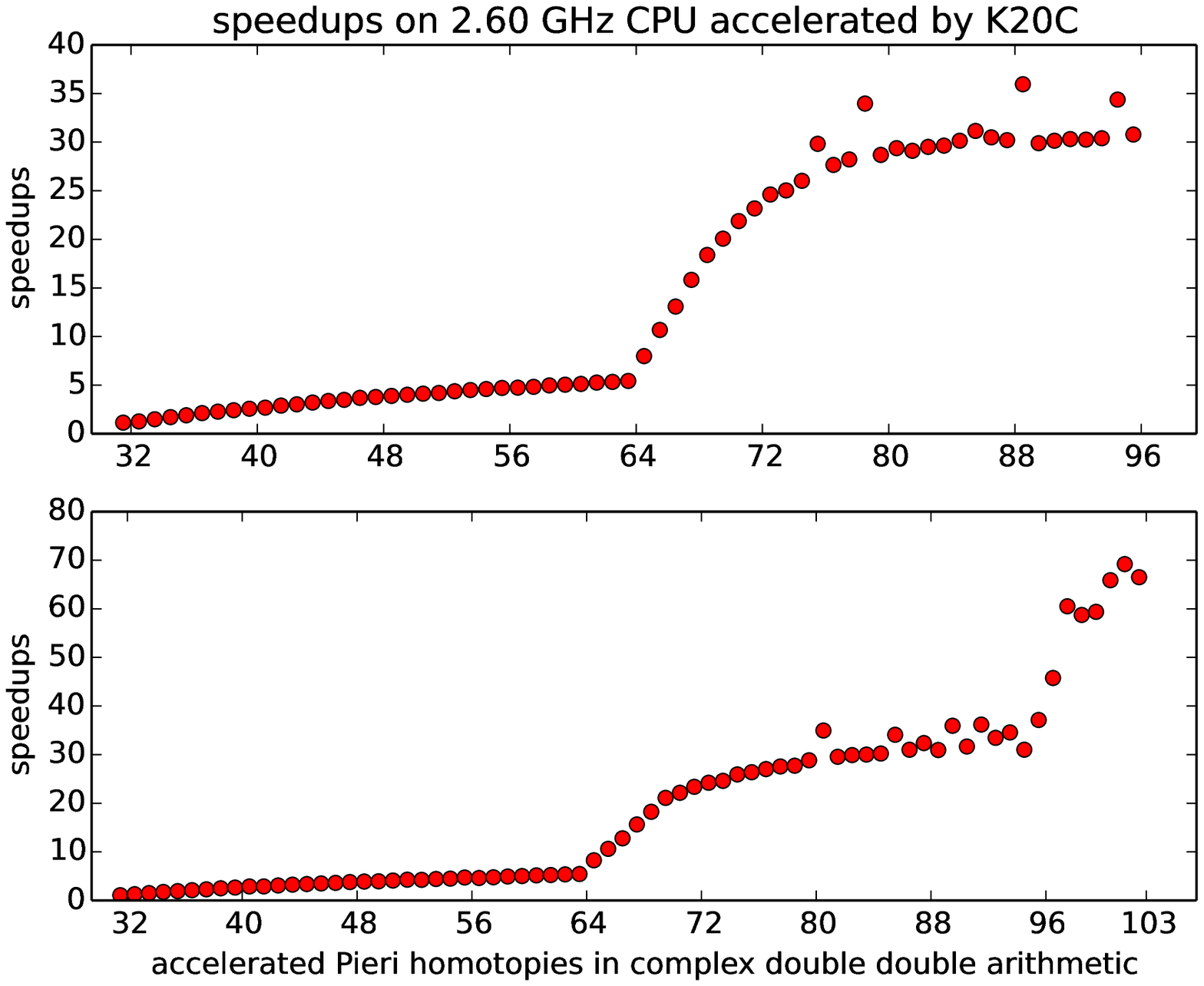,width=8.3cm}
\caption{Visualization of the speedups of Table~\ref{tabpieri1} on top,
with speedups of Table~\ref{tabpieri2} below.  Observe the different ranges 
of the vertical axes.}
\label{figpierispeedups}
\end{center}
\end{figure}

Double digit speedups arise after dimension 65.
After dimension 97, the speedup then almost doubles,
see Figure~\ref{figpierispeedups}.

\subsection{Running One Path of Cyclic $n$-roots}

Table~\ref{tabruncyclic} summarizes the computational results from
running one path on a homotopy to apply the monodromy on
the cyclic $n$-roots problem.
The first case where double precision does not suffice is
in dimension $n=64$, but the path can then be tracked
successfully in double double precision.
For $n = 144$, both double and double double precision
are insufficient and quad double precision is needed.

The difficulties could be explained by the higher degree
of the solution set.  Cyclic 256-roots remains a challenge.
The double digit speedups obtained by acceleration
implies that we can offset the cost of one extra level
of higher precision.

\begin{table*}[hbt]
\begin{center}
\setlength{\tabcolsep}{4pt}
\begin{tabular}{r|rrrrr|rrrrr|rrrrr}
    & \multicolumn{5}{c|}{complex double} 
    & \multicolumn{5}{c|}{complex double double}
    & \multicolumn{5}{c}{complex quad double} \\ 
$n$ & $s$  &  $m$ & cpu & gpu & {\tt S}~~
    & $s$  &  $m$ & cpu & gpu & {\tt S}~~
    & $s$  &  $m$ & cpu & gpu & {\tt S}~~ \\ \hline
 16 & 1 &   32 &    0.00 &  0.03 &  0.14 & 1 &   20 &    0.04 &  0.06 &  0.65 & 1 &   32 &     0.48 &   0.52 &  0.92 \\ 
 32 & 1 &  100 &    0.06 &  0.16 &  0.35 & 1 &   79 &    1.03 &  0.41 &  2.53 & 1 &  100 &    12.66 &   3.62 &  3.50 \\ 
 48 & 1 &  103 &    0.17 &  0.24 &  0.72 & 1 &   78 &    3.23 &  0.61 &  5.29 & 1 &  103 &    39.46 &   5.39 &  7.32 \\ 
 64 & 0 &  987 &    4.48 &  4.15 &  1.08 & 1 &  225 &   22.94 &  2.57 &  8.92 & 1 &  987 &   229.99 &  17.93 & 12.83 \\ 
 80 & 1 &   99 &    0.73 &  0.42 &  1.74 & 1 &   75 &   14.96 &  1.15 & 13.01 & 1 &   99 &   180.37 &  10.13 & 17.81 \\ 
 96 & 1 &   95 &    1.23 &  0.52 &  2.36 & 1 &   69 &   23.17 &  1.34 & 17.26 & 1 &   95 &   289.38 &  12.64 & 22.90 \\ 
112 & 1 &  171 &    3.42 &  1.17 &  2.92 & 1 &  121 &   68.07 &  2.98 & 22.86 & 1 &  171 &   813.91 &  28.36 & 28.70 \\ 
128 & 1 &  162 &    5.66 &  1.47 &  3.85 & 1 &  123 &  102.94 &  3.88 & 26.54 & 1 &  162 &  1253.82 &  37.75 & 33.21 \\ 
144 & 0 &  214 &   12.58 &  2.67 &  4.72 & 0 & 1500 & 1487.86 & 61.59 & 24.16 & 1 &  214 & 15898.67 & 479.18 & 33.18 \\ 
160 & 1 &   68 &    4.84 &  0.87 &  5.53 & 1 &   49 &   83.11 &  2.84 & 29.31 & 1 &   68 &   998.43 &  23.96 & 41.67 \\ 
176 & 1 &  160 &   15.65 &  2.52 &  6.21 & 1 &  118 &  259.80 &  8.06 & 32.24 & 1 &  160 &  3179.81 &  70.58 & 45.05 \\ 
192 & 0 &  246 &   30.92 &  9.27 &  3.34 & 1 &  150 &  419.16 & 13.03 & 32.16 & 1 &  246 &  5054.70 & 105.69 & 47.83 \\ 
208 & 1 &  231 &   39.51 &  5.22 &  7.57 & 1 &  168 &  628.46 & 16.33 & 38.48 & 1 &  231 &  7529.02 & 147.09 & 51.19 \\ 
224 & 1 &   96 &   19.39 &  2.46 &  7.88 & 1 &   71 &  319.27 &  7.88 & 40.54 & 1 &   96 &  3925.33 &  73.76 & 53.22 \\ 
240 & 1 &  140 &   34.04 &  4.04 &  8.42 & 1 &   96 &  531.01 & 12.49 & 42.50 & 1 &  140 &  6714.01 & 119.86 & 56.01 \\ 
256 & 0 &    0 &    0.00 &  1.00 &  0.00 & 0 &    0 &    0.00 &  1.00 &  0.00 & 0 &    0 &     0.00 &   1.00 &  0.00 \\ 
272 & 1 &  160 &   58.19 &  7.19 &  8.09 & 1 &  118 &  914.24 & 19.12 & 47.82 & 1 &  160 & 10829.36 & 183.12 & 59.14 \\ 
288 & 0 &    0 &    0.00 &  1.00 &  0.00 & 0 &    0 &    0.00 &  1.00 &  0.00 & 0 &    0 &     0.00 &   1.00 &  0.00 \\ 
304 & 1 &  142 &   81.04 &  8.05 & 10.07 & 1 &  103 & 1176.29 & 22.87 & 51.44 & 1 &  142 & 13992.60 & 226.78 & 61.70 \\ 
320 & 0 &    0 &    0.00 &  1.00 &  0.00 & 0 &    0 &    0.00 &  1.00 &  0.00 & 0 &    0 &     0.00 &   1.00 &  0.00 \\ 
336 & 1 &  157 &  105.30 & 11.12 &  9.47 & 1 &  114 & 1772.97 & 33.26 & 53.31 & 1 &  157 & 20807.27 & 327.25 & 63.58 \\ 
352 & 1 &  121 &   93.89 &  9.78 &  9.60 & 1 &   90 & 1621.15 & 28.75 & 56.39 & 1 &  121 & 18881.13 & 290.36 & 65.03 \\ 
\end{tabular}
\caption{Running one path of cyclic $n$-roots in various
precisions and in various dimensions.
Data in the column under the header~$s$ indicates
success by~${\rm 1}$ or failure by~${\rm 0}$.
The number of predictor-corrector steps equals~$m$.
The units of the times on CPU and GPU are seconds.
The last column for each dimension and precision
contains the speedup~{\tt S}.}
\label{tabruncyclic}
\end{center}
\end{table*}

\begin{figure}[hbt]
\begin{center}
\epsfig{figure=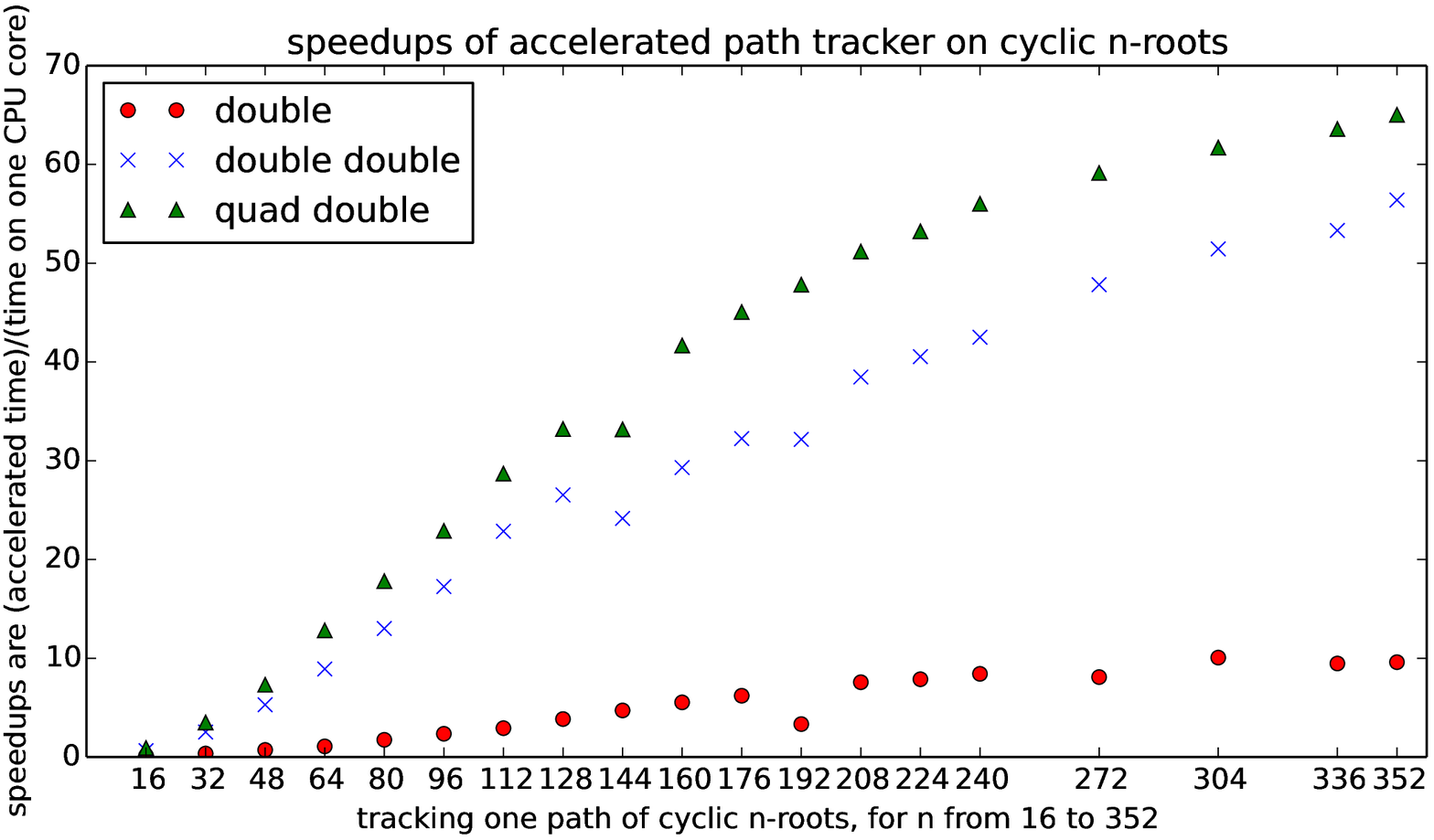,width=8.3cm}
\caption{Visualization of the speedups of Table~\ref{tabruncyclic}.
In double precision, the dimensions are too small to achieve double
digit speedups, but then also the precision may be insufficient.
In double double precision, speedups are good as soon as the dimension
reaches one hundred and they improved even further in quad double
precision.}
\label{figcyclicspeedups}
\end{center}
\end{figure}

The data in Table~\ref{tabruncyclic} for complex double
and complex double double precision is visualized in 
Figure~\ref{figcyclicspeedups}.

Concerning the data in Table~\ref{tabruncyclic},
let us compare the accelerated times in double double precision
to the times on one CPU core in double precision.
For the last line, observe that it takes 93.89 seconds
to track one path in double precision without acceleration.
With acceleration tracking one path in double double
precision takes 28.75 seconds, so we
can double the precision and still be three times faster
than in double precision without acceleration.
Speedups computed in Table~\ref{tabruncyclic} 
are shown in Figure~\ref{figcyclicspeedups}.
We see that in double double precision, the speedups rise
faster as the dimension increases than in double precision.

\section{Concluding Remarks}

This paper provides a description for an accelerated path tracker
for polynomial homotopies.  On two classes of benchmark problems
we illustrate that for sufficiently high dimensions with acceleration
we can compensate for the extra cost of high precision arithmetic.

\section*{Acknowledgments} 

This material is based upon work supported 
by the National Science Foundation under Grant No.\ 1440534.
The Microway workstation with the NVIDIA Tesla K20C
was purchased through a UIC LAS Science award.
We thank the reviewers for their comments and suggestions
which improved the quality of this paper.

\bibliographystyle{plain}

\end{document}